\documentclass[sigconf]{acmart}
\usepackage{graphicx,pifont}
\usepackage{threeparttable}
\usepackage{multirow}
\usepackage{algorithm}
\usepackage{algorithmicx}
\usepackage{algpseudocode}
\usepackage{color}
\usepackage{comment}
\usepackage{geometry}
\renewcommand\footnotetextcopyrightpermission[1]{}

\AtBeginDocument{%
  \providecommand\BibTeX{{%
    \normalfont B\kern-0.5em{\scshape i\kern-0.25em b}\kern-0.8em\TeX}}}

\acmConference[DAC '2024]{Design Automation Conference}{June 23--27, 2024}{San Francisco, CA}

\begin{document}
\newcommand{\cmark}{\ding{51}}%
\newcommand{\xmark}{\ding{55}}%
\title{Older and Wiser: The Marriage of Device Aging and Intellectual Property Protection of Deep Neural Networks}

\author{\fontsize{10}{8}\selectfont Ning Lin$^{1,2}$, Shaocong Wang$^{1,2}$, Yue Zhang$^{1,2}$, Yangu He$^{1,2}$, Kwunhang Wong$^{1,2}$, Arindam Basu$^5$, \\ Dashan Shang$^4$, Xiaoming Chen$^{3,*}$ and Zhongrui Wang$^{1,2,*}$}
\renewcommand{\authors}{N. Lin, S. Wang, Y. Zhang, Y. He, K. Wong, A. Basu, D. Shang, X. Chen, Z. Wang}

\affiliation{%
  \institution{\fontsize{8}{8}\selectfont$^1$Department of Electrical and Electronic Engineering, The University of Hong Kong, Hong Kong, China; $^2$ACCESS – AI Chip Center for Emerging Smart Systems, InnoHK Centers, Hong Kong Science Park, Hong Kong, China; $^3$Institute of Computing Technology, Chinese Academy of Sciences, Beijing, China; $^4$Key Laboratory of Microelectronics Devices and Integrated Technology, Institute of Microelectronics, Chinese Academy of Sciences, Beijing, China; $^5$Department of Electrical Engineering, City University of Hong Kong, Hong Kong, China.
  \\ $^*$Corresponding authors: chenxiaoming@ict.ac.cn; zrwang@eee.hku.hk}
  \country{}
}

\renewcommand{\shortauthors}{N. Lin et al.}

\begin{abstract}
Deep neural networks (DNNs), such as the widely-used GPT-3 with billions of parameters, are often kept secret due to high training costs and privacy concerns surrounding the data used to train them. Previous approaches to securing DNNs typically require expensive circuit redesign, resulting in additional overheads such as increased area, energy consumption, and latency. To address these issues, we propose a novel hardware-software co-design approach for DNN intellectual property (IP) protection that capitalizes on the inherent aging characteristics of circuits and a novel differential orientation fine-tuning (DOFT) to ensure effective protection.

Hardware-wise, we employ random aging to produce authorized chips. This process circumvents the need for chip redesign, thereby eliminating any additional hardware overhead during the inference procedure of DNNs. Moreover, the authorized chips demonstrate a considerable disparity in DNN inference performance when compared to unauthorized chips. Software-wise, we propose a novel DOFT, which allows pre-trained DNNs to maintain their original accuracy on authorized chips with minimal fine-tuning, while the model's performance on unauthorized chips is reduced to random guessing. Extensive experiments on various models, including MLP, VGG, ResNet, Mixer, and SwinTransformer, with lightweight binary and practical multi-bit weights demonstrate that the proposed method achieves effective IP protection, with only 10\% accuracy on unauthorized chips, while preserving nearly the original accuracy on authorized ones.
\end{abstract}








\maketitle
\section{Introduction}

In recent years, foundation models like GPT-3~\cite{brown2020language}, have driven the advancement of artificial intelligence to unprecedented levels. However, the security concerns associated with these models are garnering increasing attention. This is primarily due to the fact that well-trained deep neural networks (DNNs) represent valuable intellectual property (IP) assets, as they necessitate significant investments in extensive datasets, cutting-edge hardware, and skilled professionals to design the architecture and optimize hyperparameters. While these factors contribute to the commercial profitability of DNNs, they also introduce security vulnerabilities. For instance, the leakage of confidential parameters, such as weights, could enable attackers to develop Trojan horses or adversarial examples, thereby jeopardizing the proper functioning of DNNs.

Existing protection methods can be classified into three categories: watermarking, encryption and circuit obfuscation. However, these methods either provide only a verification capability or incur additional performance overhead during the inference phase. 

Watermarking-based protection, encompassing both white-box and black-box techniques (e.g., \cite{uchida2017embedding,guo2018watermarking,sheybani2023zkrownn}), primarily operates at the software level. As an example of a typical black-box method, during the training phase, owners introduce distinct marks and labels to a minor subset of the training dataset~\cite{guo2018watermarking}. In the inference stage, the DNN output on alternative platforms is authenticated by incorporating a specific mark into the input data. If the DNN's output aligns with the expected result, the validator establishes ownership. Although its additional  hardware overhead is almost negligible, watermarking cannot prevent model stealers from normally using DNNs for free. 

\begin{figure}[!tbp]
\centerline{\includegraphics[width=1.0\columnwidth]{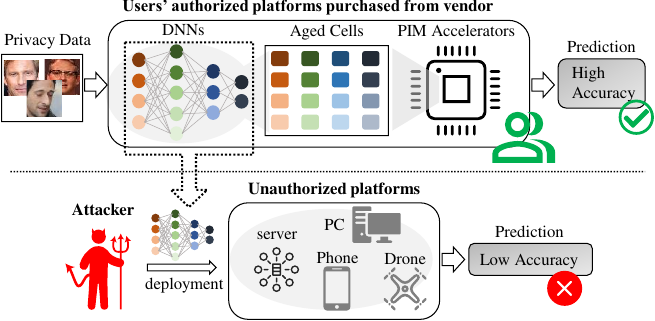}}
   \vspace{-10pt}  
\caption{IP protection overview.}
\label{fig:over}
   \vspace{-14pt}  
\end{figure}
Encryption protection involves using particular encryption algorithms to encrypt the weights of DNNs. For instance,  Zuo \textit{et al.} \cite{zuo2021sealing} propose a smart encryption scheme that stores AES-encrypted weights of DNNs in the main memory and decrypts them when they are transferred to the processing units. 
Nevertheless, the mathematically complex  AES-based encryption methods tend to consume a significant amount of area overhead and power consumption.
To make protection lightweight, Huang \textit{et al.} \cite{huang2020xor} implemented XOR encryption by modifying the 6T-SRAM cell with dual wordlines and corresponding peripheral circuits to protect DNNs running on SRAM-based accelerators. 
Cai \textit{et al.} \cite{cai2019enabling} proposed a sparse fast gradient encryption to protect DNNs running on RRAM-based accelerators by encrypting a small proportion of weights. 

Circuit obfuscation protection requires modifying the circuit structure according to specific operational logic to achieve normal computation. Chakraborty \textit{et al.} \cite{chakraborty2020hardware} proposed a simple neuron locking scheme by modifying the multiply and accumulate units of TPU-like accelerators, which combines a hardware root-of-trust (i.e., secret key embedded on-chip) to protect DNNs. 
Zhao \textit{et al.} \cite{zhao2022sra} used a stochastic computing-based scheme and custom hardware accelerator architecture to protect the weights of DNNs running on RRAM crossbars. 

This raises the question of whether it is possible to secure DNN workloads without changing the physical architecture of the accelerators. Aging is commonly observed on almost all circuits, which typically been seen as a drawback on hardware's performance. However, when protecting the intellectual property of DNNs, this flaw can be turned into an advantage as shown in Fig.~\ref{fig:over}. We introduce several innovations to achieve this.
\begin{itemize}
\item We propose a novel hardware-software co-design to protect the IP of DNNs without any circuit modification, resulting in zero additional latency, power, and area overheads during the inference phase of DNNs.
\item Hardware-wise, we leverage the inherent aging mechanism of transistors to create authorized chips with distinct operational mechanisms compared to unauthorized hardware platforms. Our approach enhances the security of the model by randomly selecting the aging configurations.
\item Software-wise, we introduce a differential orientation fine-tuning method that enables DNNs to achieve high accuracy on authorized chips and low accuracy on unauthorized chips with only few fine-tuning procedures.
\item Various models, comprising MLP, VGG, ResNet, Mixer, and SwinTransformer, have effectively substantiated the efficacy of our method. Furthermore, the utility and security have also been rigorously appraised.
\end{itemize}

\begin{figure}[!tbp]
\centerline{\includegraphics[width=1.0\columnwidth]{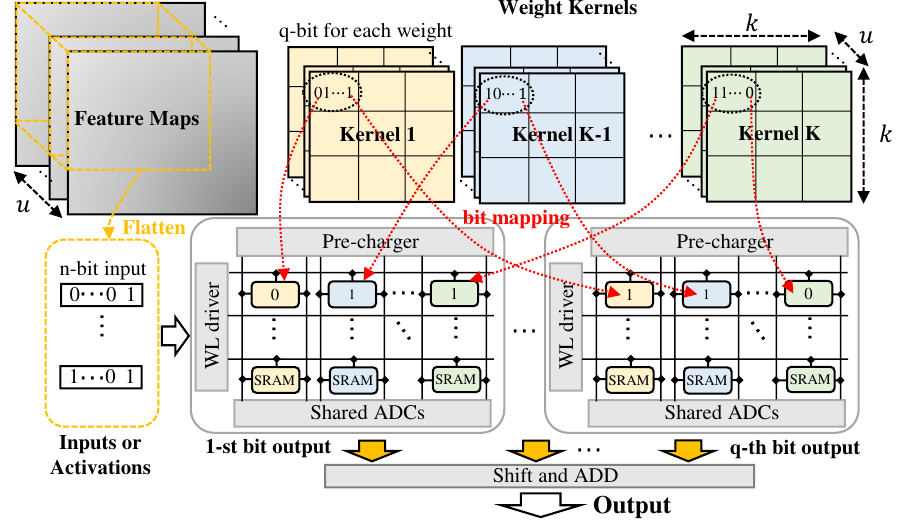}}
   \vspace{-10pt}  
\caption{Principle of SRAM-based PIM accelerators.}
\label{fig:sram}
   \vspace{-14pt}  
\end{figure}

\section{Preliminaries and Motivation}
\subsection{SRAM-based DNN Accelerators}
In the last decade, a large number of hardware DNN accelerators have been developed to improve both throughput and energy efficiency.
Traditional von Neumann architecture accelerators (e.g.,~\cite{TPU2017ISCA}) suffer from large energy and time overheads due to data shuttling  between the main memory and computing components. Process-in-memory (PIM) accelerators that feature collocation of memory and processing units have been proposed to mitigate the von Neumann bottleneck. Representative PIM-based accelerators include RRAM- and SRAM-based DNN accelerators. In this paper, we choose SRAM-based accelerators for faster programming speed and mature technology and proven reliability.

Fig.~\ref{fig:sram} depicts the basic principle of SRAM-based DNN accelerators~\cite{6TSRAM2018DAC}, which are mainly composed of a set of SRAM arrays. Well-trained weights are stored in the SRAM cells. Each sliding window on the input feature maps is flattened into a vector, which is converted into word line (WL) voltages using the WL driver. The SRAM arrays perform matrix-vector multiplications between the weights stored in the arrays and the input vectors. 
$q$-bit weights require $q$ SRAM arrays to store different bits of weights. A $q$-bit weight $w^l(q)$ in layer $l$ can be represented by 
\begin{equation}
w^{l}(q)=2^{q-1}w^{l}[q]+\cdots +2^{1}w^{l}[2] +2^{0}w^{l}[1],
\label{eq:wq}
\end{equation}
where $w^{l}[i]$ is a binary weight bit and $i$ is the bit index ($i \in \{1,2,\cdots,q\}$), and the $i$-th SRAM array stores the $i$-th bits of weights of different kernels in the $l$-th layer of DNNs. An $n$-bit input can also be expressed in a similar way to Eq.~\eqref{eq:wq}. These input vectors are streamed in through a bit-serial manner, and the same index bits of inputs are supplied at the same time. 

\begin{figure}[!t]%
    \centering
    {\includegraphics[width=1.0\columnwidth]{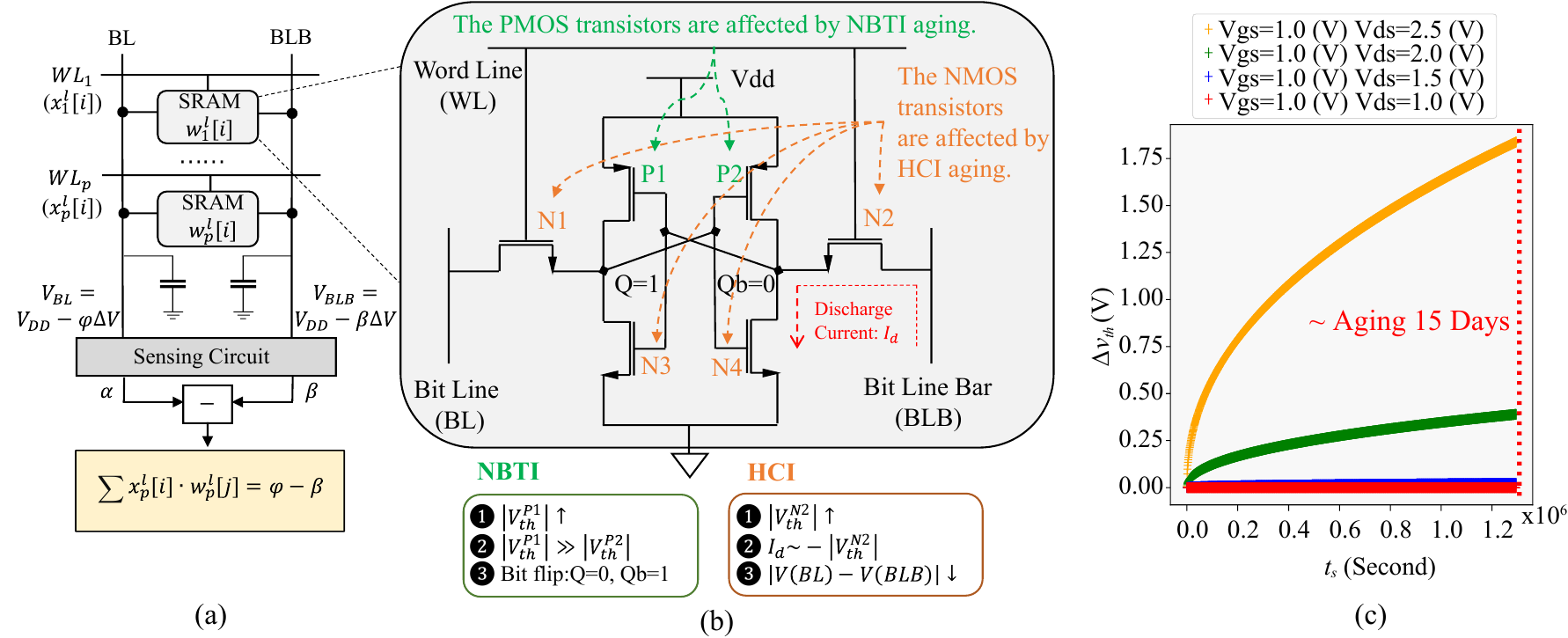} } \\
       \vspace{-10pt}  
    \caption{Dot-product computation (a) and aging mechanism in a 6T-SRAM cell (b). Threshold voltage changes due to the HCI-induced Aging (c).}%
    \label{fig:6tsram}%
       \vspace{-14pt}  
\end{figure}

Fig.~\ref{fig:6tsram} (a) illustrates the in-memory computation principle of an SRAM column, which computes an inner product between an input bit vector $\left[x_{1}^{l}[i],x_{2}^{l}[i],\cdots,x_{p}^{l}[i]\right]$ ($p=k\times k\times u$) and the stored weight bit vector $\left[ w_1^{l}[j],w_{2}^{l}[j],\cdots,w_{p}^{l}[j]\right]$, where $i$ and $j$ are the bit indexes of inputs and weights, respectively. Though the involved input bits and weight bits are both digital, the inner product is computed in the analog domain. The inner product of the two vectors is manifested by the voltage difference between BL and BLB of the column, which is proportional to $\sum_{p} x_{p}^{l}[i]\cdot w_{p}^l[j]$. The voltage difference between the bit line (BL) and bit line bar (BLB), which is called \textit{read voltage difference} in this work, is sensed and digitized by an analog-to-digital converter (ADC). 

\subsection{Device Aging}

Hot Carrier Injection (HCI) and Negative Bias Temperature Instability (NBTI) are two primary factors contributing to the aging of NMOS and PMOS transistors, respectively~\cite{reliabilitybook,WTDMR,WyISQED,survey}. HCI and NBTI both cause an increase in the threshold voltage ($v_{th}$) of transistors, which can result in the malfunctioning of SRAM cells, 
\begin{equation}
\Delta v_{th} \propto  \zeta \cdot t_s^{\chi },
\end{equation} 
where $t_s$ is the stress time,  ${\zeta}$ is a coefficient that depends on the aging type and working environment of the circuit, and ${\chi}$ is a positive exponent which differs from the aging type. The detailed formulas of NBTI- and HCI-induced aging can be found in ~\cite{WTDMR,reliabilitybook,WyISQED,survey}.
Fig.~\ref{fig:6tsram} (c) illustrates the trend of threshold voltage growth with stress time at different supply voltages (i.e., $V_{ds}$) for HCI-induced aging under the 45nm technology node. The increase in $v_{th}$ is faster with a higher supply voltage applied to an NMOS transistor. When $V_{ds}=2.0$V is applied to the NMOS transistor for about 15 days, $v_{th}$ increases by about 0.5V. With the power supply voltage $V_{ds}=2.5$V, $v_{th}$ increases by about 2.3V. Therefore, changing the working conditions can rapidly age the chip. 
Even though both types of aging can increase the threshold voltage, they result in different behaviors of aged SRAM devices.


\textbf{NBTI can cause read bit flip.}  
Fig.~\ref{fig:6tsram} (b) illustrates the NBTI aging impact on a 6T-SRAM cell. If the SRAM has been configured to Q=1 and Qb=0 for a sufficiently long time, the threshold voltage of the P1 transistor ($v^{P1}_{th}$) will experience a clear increase due to the influence of NBTI. When $| v^{P1}_{th}| \gg  | v^{P2}_{th}|$, bit flip may occur, that is, Q=0 and Qb=1. Under this circumstance, the read result of the SRAM will flip as well.

\begin{figure}[!tbp]
\centerline{\includegraphics[width=1.0\columnwidth]{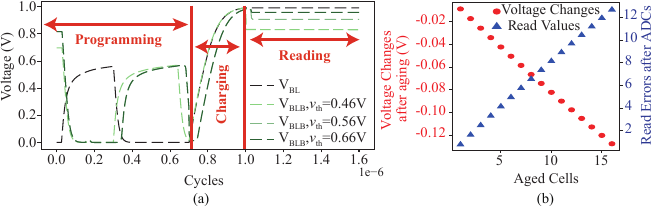}}
\vspace{-10pt}
\caption{Voltage changes of 6T-SRAM cells (a) and read errors after ADC (b).}
\label{fig:faged}
\end{figure}

\textbf{HCI can change the SRAM read voltage.} 
As illustrated in Fig.~\ref{fig:6tsram} (b), when the WL is at logic high, the NMOS transistors N1 and N2 discharge BL and BLB, respectively. While transistor N4 is on and transistor N3 is off, the charge on the BL remains constant and the BLB discharges. However, if the discharge current flowing through N2 is reduced due to HCI-induced aging, the threshold voltage of N2 increases. As a result, the read voltage difference (i.e., $\Delta V_{read}$) of the SRAM cell decreases, which can lead to changes in the layer outputs of DNNs, and associated accuracy degradation.
\subsection{Motivation}
NBTI in SRAMs can cause bit flip, which causes ``digital'' errors and can easily be detected as defective chips.  On the contrary, HCI impacts SRAMs in an ``analog'' way and cannot be detected through the I/O interface of chips. Indeed, it is reasonable to assume that only the I/O interface can be accessed when chips are sold. Thus, HCI-induced aging is more concealed that can prevent attackers from obtaining specific aging information of SRAM-based DNN accelerators through I/O interface.  From this point of view, we utilize HCI-induced aging to protect the security of DNNs' weights.

\section{Marriage of aging and IP protection}
\subsection{Threat Models}
Once the attacker obtains the weights of DNNs, (s)he could potentially clone the DNNs to another chip. As a result, attackers may gain financial interests by reselling high performance DNNs without authorization from the DNN providers.

\textbf{Attacker's Capabilities.} End users may require a new DNN when facing with new tasks, so the DNN vendor needs to transmit it through the internet, which provides the possibility for attackers to steal these DNNs by the man-in-the-middle attacks. DNNs running on accelerators may also be stolen by attackers via side-channel attacks \cite{hua2018reverse}.

\textbf{Attacker’s Limitations.} Consistent with literatures (e.g.,~\cite{lin2020chaotic,guo2018puf}), we assume attackers do not have the private data to retrain the stolen parameters of DNNs, as these data are usually private and precious. Also, data protection laws and commercial values protect these data from being made public. 

\subsection{HCI Aging Impact}

We first analyze how HCI impacts the accuracy of DNNs running on SRAM-based PIM accelerators. We utilize LTspice to simulate the BL (BLB) voltage comparison of a 16$\times$1 SRAM array with different HCI-induced aged cells under 45nm technology node in Fig.~\ref{fig:faged} (a). When the SRAM stores `1' (where Q=1 and Qb=0) and the read voltage signal is at high level, NMOS transistors N2 and N4 are turned on, forming a discharge path of BLB as shown in Fig.~\ref{fig:6tsram} (b). At this point, the voltage of $V_{BLB}$ is $V_{BLB}=V_{pre}-\Delta$. Conversely, since transistor N3 is turned off, the voltage of BL remains unchanged as $V_{BL}=V_{pre}$. As the SRAM cell ages, the voltage drop of BLB changes accordingly.

Specifically, when no aging has occurred (i.e., $v_{th}=0.46$V), the read voltage of BLB (see $V_{BLB}$ in Fig.~\ref{fig:faged} (a)) is approximately 0.83V. When HCI aging occurs, the read voltage of BLB is about 0.9V when $v_{th}$ = 0.56V. As the threshold voltage increases to 0.66V, the read voltage of BLB also increases to 0.95V. Thus, the change in threshold voltage $v_{th}$ causes a corresponding change in the BLB voltage, which in turn results in read errors. Fig.~\ref{fig:faged} (b) shows that as the number of aged SRAM cells in a population increases, the read voltage of the aged circuit becomes smaller than that of the unaged case. This introduces read errors after digitization by ADCs.
For instance, if 15 SRAM cells that all store `1' connected to the same BL and BLB are aged, although the voltage difference after aging ($v_{th}=0.66$V) is reduced by about only 0.12V, a read error value of `12' will be generated after ADCs with 0.01V read voltage interval.
Hence, we can utilize the HCI aging mechanism to develop unique DNN accelerators.
This approach is fundamentally distinct from traditional accelerator protection. Investigating the impact of aged chips on DNNs' accuracy becomes a critical issue, which will be explored in the following contents.

\begin{figure}[!tbp]
\centerline{\includegraphics[width=0.9\columnwidth]{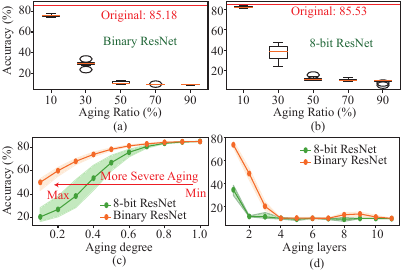}}
\vspace{-10pt}
\caption{Aging ratio impact on binary (a) and 8-bit ResNet (b). Aging degree impact (c) and layer impact (d). Results are obtained after 100 trials.}
\label{fig:fadg}
\vspace{-12pt}
\end{figure}

\textbf{\ding{182} Aging Ratio.}
Assuming that the aged SRAM cells are distributed in a two-dimensional aging matrix $\textbf{M}_{aging}$, consisting of aged element $m_{aging}$ and normal element $m$, then the aging ratio $\sigma$ can be expressed as 
\begin{equation}
\sigma=\frac{Number(m_{aging})}{Number(m_{aging})+Number(m)} \times 100\%. 
\end{equation}

To investigate the impact of aging ratio, we evaluate the accuracy of ResNet on CIFAR10 under different aging ratios, as shown in Fig.~\ref{fig:fadg} (a) and (b).  The aging degree is 0.24 (aging degree is defined in the following sub-section).  Obviously, the accuracy decreases as the aging ratio increases. For instance, the accuracy of the binary (8-bit) ResNet decreases by an average of 50\% (40\%) on aged SRAM accelerators when the aging ratio is 30\%. As the aging ratio increases to 50\%, the accuracy drops to about 10\%.

\textbf{\ding{183} Aging Degree.}
Aging degree refers to the read voltage difference ratio  between the aged read voltage difference $\Delta V^{Aged}_{BL, BLB}$ and the unaged read voltage difference $\Delta V^{Unaged}_{BL, BLB}$ of an SRAM cell, which is defined as 
\begin{equation}
\alpha = \Delta V^{Aged}_{BL, BLB}/\Delta V^{Unaged}_{BL, BLB}, 
\label{eq:dg}
\end{equation}
where $0<\alpha \leq 1$. Here $\alpha=1$ indicates that no aging has occurred. The lower the aging degree $\alpha$, the more severe the aging is. As the aging degree $\alpha$ decreases, the read voltage of the aged device also decreases, which leads to a drop in DNN accuracy. Fig.~\ref{fig:fadg} (c) shows the accuracy of binary and 8-bit ResNet under different aging degrees with aging ratio $\sigma$ equals to 30\%. The accuracy drops by approximately 8\% (12\%) when the aging degree is 0.6, and it sharply drops to about 61\% (27\%) for the binary (8-bit) ResNet when the aging degree decreases to 0.2.

\textbf{\ding{184} Aging Layers Impact.}
Fig.~\ref{fig:fadg} (d) shows that as the number of aging layers (where aging ratio and aging degree equal to 90\% and 0.24, respectively) increases, the accuracy drops dramatically. For example, when randomly choosing four layers are aged in a binary ResNet, its accuracy drops close to 10\%, and only two layers need to be aged for an 8-bit ResNet to completely lose its classification ability.
Therefore, the inference accuracy on aged chips can dramatically differ from that of unaged chips, which motivates the development of DNNs that perform well on aged chips but poorly on unaged chips.
\subsection{HCI-Based Authorized Accelerators}
\textbf{Step-I: Randomly Aging Generation.}
For each layer $l\in \{1,2,\cdots,L\}$ of a DNN, we first randomly generates $q$ aging matrices, $\textbf{M}^{l}[i]$ for each weight bit index $i$ ($i \in \{1,2,\cdots,q\}$), with the same size as the weight bit matrix $\textbf{W}^{l}[i]$ in the $l$-th layer mapped on one or multiple SRAM arrays according their capacity limitation. 

Assume that for a single SRAM cell, the read voltage differences corresponding to `+1' and `-1' are $\Delta V$ and $-\Delta V$, respectively. After aging with a degree of $\alpha$ (where $0<\alpha \leq 1$, see Eq.~\eqref{eq:dg}), the corresponding read voltage differences become $\alpha \Delta V$ and $-\alpha \Delta V$, respectively. The aging matrix $\textbf{M}^{l}[i]$ consists of the following values: $m^{l}[i]$ for unaged cells and ${m}^{l}_{aging}[i]$ for aged cells, mathematically 
\begin{equation}
 m^{l}[i]= \Delta V,  \;  m^{l}_{aging}[i] = {\alpha}m^{l}[i]. 
\end{equation}

The positions of the aged cells, aging ratio and aging degree in the aging matrices are randomly generated.

State-of-the-art DNNs typically have a large number of layers, and thus, SRAM computing resources may not be sufficient to hold all layers on different SRAM arrays. In this case, the same SRAM arrays may be reused, and different layers will share the same aging matrices. We use the layer with the most number of weights to generate a prototype aging mask $\textbf{M}^{pro}[i]$ for each bit index, and the other layers use subsets of it, which is  mathematically 
\begin{equation}\label{eq:reuse}
 \textbf{M}^{l}[i] \subseteq \textbf{M}^{pro}[i]. 
\end{equation}


\textbf{Step-II: Differential Orientation Fine-tuning (DOFT).} 
After obtaining the aging matrices by reading the real SRAM-based PIM chip, the fine-tuning procedure is performed on independent hardware (e.g., GPU/TPU servers) rather than on the actually aged chips due to the challenges involved in developing additional hardware and software drivers to shuttle inference results and updated weights between servers and deliberate-aged chips. Moreover, the procedure of programming updated weights into SRAM cells during training incurs significant latency overhead. Therefore, to overcome these challenges, an aged chip simulation is developed to reflect the aging situation during fine-tuning.

\textbf{\underline{Authorized Chip Simulation.}} 
Authorized chips mean the chips with deliberate aging. For weights represented by $q$ bits (each weight bit is either `+1' or `-1' to match the electrical properties of 6T SRAMs), the matrix-vector multiplication of weights and inputs $A^{l-1}$ of layer $l$ can be expressed as 
\begin{equation}
\begin{aligned}
\!\!O^{l} &\!=\! (2^{q-1}\!\times\!\textbf{M}^{l}[q]\odot \textbf{W}^{l}[q] 
 + \cdots+  2^{1}\times\textbf{M}^{l}[2]\odot \textbf{W}^{l}[2]\\
 &\quad   + 2^{0}\times\textbf{M}^{l}[1]\odot \textbf{W}^{l}[1] )\times A^{l-1}  \vspace{-4pt}
\label{eq:IN-forward-mac},
\end{aligned} 
\end{equation}
where $\odot$ is the Hadamard Product operation, and $\textbf{W}^{l}[i]$ and $\textbf{M}^{l}[i]$ are the weight matrix and the corresponding aging matrix of bit index $i$ ($i \in \{1,2,\cdots,q\}$), respectively. Then the output of layer $l$ on after ADCs can be represented by 
\begin{equation}
A^{l} \gets h\left( \left \lfloor   O^{l} / \Delta V^{inter} \right \rfloor  \right)  
\label{eq:adc}, 
\end{equation}
where $\Delta V^{inter}$ is the read voltage interval of ADCs. 
Eq.~\eqref{eq:adc} simulates the function of nonlinear module and ADCs in SRAM-based PIM DNN accelerators. $h(\cdot)$ is the nonlinear activation function (e.g., relu or sigmoid), which is realized by digital modules in the peripheral circuits.

\textbf{\underline{Unauthorized Chip Simulation.}} The term ``unauthorized'' is used to describe chips that have not undergone the deliberate aging treatment. In practice, commonly used GPUs/TPUs, general-purpose CPUs, and DNN accelerators without deliberate aging can be considered as unauthorized chips. Eqs.~\eqref{eq:IN-forward-mac} and \eqref{eq:adc} remain valid in this case, but the elements in the aging matrices are simply composed of $\Delta V$ and $-\Delta V$, corresponding to the `+1' and `-1' values, respectively.

To distinguish between authorized and unauthorized chips, we introduce a penalty term in the loss function to reflect the accuracy gap through the objective loss function of the DOFT method, which is defined as 
\begin{equation}
\begin{aligned}
\min \mathcal{L}_{DOFT} &= \min_{X} \mathbb{E} \mathcal{L}(Y^a,Y^t) + \lambda \times \max_{X} \mathbb{E} \mathcal{L}(Y^u,Y^t) \\
&\simeq  \min_{X} ( \mathbb{E} \mathcal{L}(Y^a,Y^t) -\lambda  \times \mathbb{E} \mathcal{L}(Y^u,Y^t))
\label{eq:loss}, 
\end{aligned} 
\end{equation}
where $X$ and $Y^{t}$ represent the inputs and ground-truth labels of the entire dataset, respectively. $Y^{a}$ and $Y^{u}$ are the outputs on authorized and unauthorized chips, respectively. $\mathcal{L}$ denotes the loss function, which can be either cross-entropy loss or mean squared error loss. $\mathbb{E}$ is the operation of mathematical expectation. $\lambda$ is a hyper-parameter that controls the importance of the accuracy loss between authorized and unauthorized chips. Thus, this loss function aims to achieve two objectives. The first is to improve accuracy on authorized chips, and the second is to reduce accuracy on unauthorized chips.

\begin{figure*}[!t]%
    \centering
    {\includegraphics[width=2.1\columnwidth]{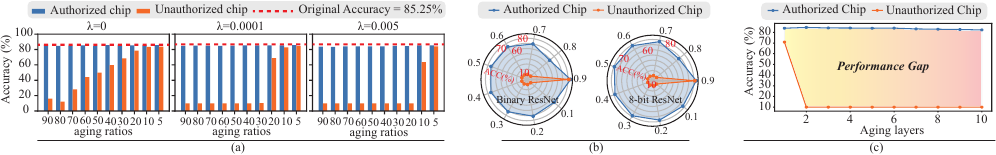} } \\
    \vspace{-10pt}  
    \caption{IP Protect effect with different aging ratios (a), aging degrees (b) and aging layers (c) on ResNet.}%
    \label{fig:acc}
\end{figure*}

To achieve the objectives, we employ the gradient backpropagation method to update the weights of DNNs during fine-tuning.  Specifically, each weight bit $w^{l}[i]$ ($i\in \{1,2,\cdots,q\}$) in each weight bit matrix $\textbf{W}^{l}[i]$ is updated by \vspace{-4pt}
\begin{equation}
w_{fp}^{l}[i] \gets  w_{fp}^{l}[i] -\eta \times\frac{\partial \mathcal{L}_{DOFT}}{\partial a^{l}}\times\frac{\partial a^{l}}{\partial o^{l}}\times\frac{\partial o^{l}}{\partial w_{fp}^{l}[i]}
\label{eq:back}, 
\end{equation}
\begin{equation}
w^{l}[i] \gets Binary(w_{fp}^{l}[i])
\label{eq:back-binary}, 
\end{equation}
where $\eta$ is the learning rate, and $a^l$ and $o^l$ are elements in the output matrix $A^{l}$ and multiplication matrix $O^{l}$, respectively. $w_{fp}^l[i]$ is full-precision intermediate weight in layer $l$ for bit index $i$ (each $w^l[i]$, $i\in \{1,2,\cdots,q\}$, has an associated full-precision intermediate weight $w_{fp}^l[i]$).
The third gradient term in Eq.~\eqref{eq:back-binary} involves the derivative binary quantization (where $Binary(\cdot)$ function in Eq.~(\ref{eq:back-binary}) quantizing $w_{fp}^l[i]$ to `+1' or `-1'.), which is non-differentiable. To address this issue, the straight-through estimator (STE)~\cite{bengio2013estimating}  is utilized to approximate its gradient by 
\begin{equation}
\partial o^{l} / \partial w_{fp}^{l}[i] \approx \partial o^{l} / \partial w^{l}[i]. 
\end{equation}

After few epochs of fine-tuning, once the loss function $\mathcal{L}_{DOFT}$ has reached a plateau and is no longer decreasing, the well-trained weight bit matrices $\textbf{W}^{l}[i]$ ($i \in \{1,2,\cdots,q\}$) can be deployed onto deliberate-aged authorized chips. 


\section{Evaluation}
\subsection{Experimental Setup }
\label{section:setup}
\textbf{Benchmarks.} The evaluation was carried out using an MLP consisting of two fully-connected layers on the MNIST dataset~\cite{MNIST}, as well as VGG~\cite{simonyan2014very}, ResNet~\cite{he2016deep}, Mixer~\cite{tolstikhin2021mlp}, and Swin Transformer~\cite{liu2021swin} on the CIFAR-10 dataset~\cite{krizhevsky2009learning}. To evaluate the versatility of the method on both edge and server-side applications, the model's weights are quantized into single-bit (binary) and multi-bit (8-bit) representations. All aged models successfully completed the proposed differential orientation fine-tuning procedure within few epochs.

\textbf{Accelerators.} To simulate SRAM-based accelerators~\cite{6TSRAM2018DAC}, we used LTSpice and derived the parameters of NMOS and PMOS transistors from the 45nm models for low-power applications in the Predictive Technology Model~\cite{PTM}. 
The SRAM array size is fixed at 64$\times$64, the ADC resolution is set to 7 bits, which generates a read voltage interval ($\Delta V^{inter}$) of 0.01V and covers output values ranging from `-64' to `+64' from the SRAM array without affecting the accuracy of the well-trained DNNs.
We developed a PyTorch based toolchain to evaluate accuracy of various DNNs running on (un)aged SRAM-based DNN accelerators according to prior architecture~\cite{6TSRAM2018DAC}.
Unless otherwise specified, the aging degree $\alpha$=0.24, and aging ratio $\sigma$=90\%. The number of aged layers encompasses all layers of the model.

\subsection{IP Protection Effect}
\textbf{Sensitivity Analysis.} Fig.~\ref{fig:acc} illustrates that our proposed method achieves significantly higher accuracy on authorized chips than on unauthorized chips with different aging ratios, aging degrees and aging layers on ResNet.
As depicted in Fig.~\ref{fig:acc} (a), the fine-tuned ResNet can achieve a significant performance difference between authorized and unauthorized chips across the majority of aging ratios. For instance, when $\lambda=0.05$ and the aging ratio is higher than 20\%, the accuracy on unauthorized chips is equivalent to random guessing, while the accuracy on authorized chips remains consistent with that of original model. Moreover, the regularization term $\lambda$ is essential for performance disparity. Specifically, when $\lambda$ is set to 0, there is minimal performance difference between authorized and unauthorized chips under low aging rates. In contrast, when $\lambda$ is assigned values of 0.0001 or 0.005, the model can generate significant performance disparities across most aging ratios.

\begin{figure}[!t]
\centerline{\includegraphics[width=1.0\columnwidth]{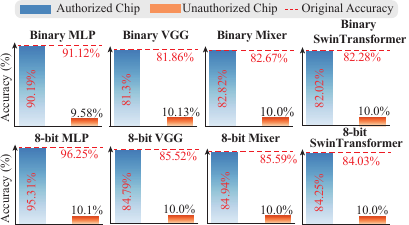}}
\vspace{-10pt}
\caption{Generalization across various DNNs.}
\label{fig:verious}
\end{figure}

We also performed a sensitivity analysis on the aging degree in Fig.~\ref{fig:acc} (b). The results demonstrate that for binary or 8-bit weight model, when the aging degree is smaller than 0.7, the accuracy running on the attacker's chip is close to random guessing, while the accuracy on our chip is guaranteed to be higher. Fig.~\ref{fig:acc} (c) demonstrates that as the number of aged layers increases, the model displays a distinct difference in accuracy  between authorized and unauthorized chips. Notably, when the number of aged layers exceeds 2, the model exhibits higher accuracy for authorized chips, while the accuracy for unauthorized chips approaches that of random guessing.

\begin{figure*}[!t]
\centerline{\includegraphics[width=2.1\columnwidth]{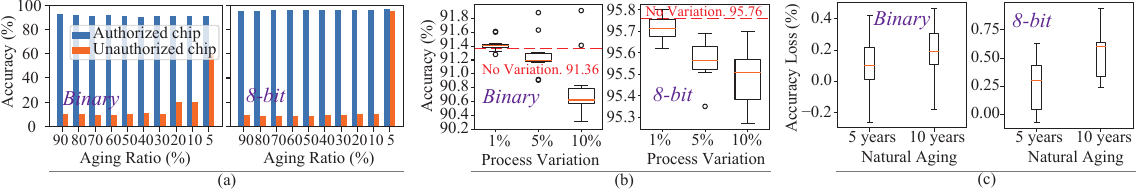}}
\vspace{-10pt}
\caption{(a) SRAM array reuse and (b) process variation analysis on MLP. (c) Natural aging evaluation on ResNet.}
\label{fig:natureaging}
\end{figure*}

\textbf{Generalization Capability.}
Fig.~\ref{fig:verious} demonstrates the effectiveness of the proposed method across various models, including lightweight MLPs, large-scale VGGs, complex Mixer structures, and Swin Transformers with attention mechanisms. The method consistently enables these models to achieve high accuracy on authorized chips, closely resembling the original model's accuracy. Simultaneously, the method ensures that the model's performance on unauthorized chip platforms is nearly equivalent to random guessing.
\subsection{Utility \& Security}
\textbf{Array Reuse.} 
For DNNs with a large number of layers or PIM accelerators with limited SRAM computing resources, it may not be possible to map all layers to the available arrays. Instead, we may need to reuse a single SRAM array for multiple layers. In such cases, the aging matrices are also shared by different layers. We can still employ the proposed DOFT method, with the only difference being that the aging matrices are shared among layers, as illustrated in Eq.~\eqref{eq:reuse}. We examine the protective effects of a total of two (binary MLP) and sixteen (8-bit MLP) binary weight matrices on a single SRAM array in Fig.~\ref{fig:natureaging} (a). For example, an aging ratio of 10\% demonstrates adequate protection. The accuracy of binary and 8-bit MLP on unauthorized attackers is approximately 20\% and 10\%, respectively. Thus, these results confirm that our method remains effective in array reuse scenarios.

\textbf{Process Variation.} There is a concern that the actual aging chip may deviate from the intended one due to process variation (PV) when the chip is deliberately aged. To evaluate the robustness, we assume that an aging ratio of $\sigma$=90\%, but the aging degree $\alpha$ in the actual aging mask is a Gaussian random variable with a mean of 0.24 and standard deviation of 1\%, 5\%, or 10\%. Fig.~\ref{fig:natureaging} (b) shows the evaluation results on binary and 8-bit MLP models with 100 trials. Our findings demonstrate that even when subjected to PV effect ranging from 1\% to 10\%, the accuracy drops by no more than 1\% point. These results suggest that the proposed method can achieve high fidelity, even when the aging matrix deviates from its intended values due to PV noise.

\textbf{Natural Aging.} 
The chip undergoes natural aging as it is used.
A natural question is whether the deliberate aging scheme can still function. We analyze it in Fig.~\ref{fig:natureaging} (c) with 100 trials. The natural aging of chips is extremely slow under normal use, as the working voltage $V_{ds}$ of SRAM-based DNN accelerators is usually small, about 0.6 to 1.2V (e.g.,~\cite{jiang2020c3sram,yin2020xnor}). From the $v_{th}$ change curve shown in Fig.~\ref{fig:6tsram} (c), it can be concluded that the smaller the $V_{ds}$, the slower the aging speed. Under the condition of $V_{ds}$=1.2V and 323.15K working temperature, the threshold voltage of NMOS transistor with $v_{th}$=0.46893V increased by only about 0.0083V(0.0114V) after 5(10) years of continuous stressing. That is to say, compared with the original unaged voltage, the threshold voltage only increases by 1.78\% to 2.43\% after five to ten years of continuous use. We performed 5-year and 10-year aging accuracy loss validation on ResNet as shown in Fig.~\ref{fig:natureaging} (c). All layers of ResNet are aged and the aging ratio is set to 90\%. Experimental results confirm that the accuracy loss in natural usage is indeed very small. In the case of 5-year or 10-year aging of the binary ResNet, the accuracy loss does not exceed 0.5\%. For the 8-bit model, the maximum accuracy loss does not exceed 1\%.

\textbf{Security Analysis.} According to Kerckhoff's Principle and Shannon's Maxim \cite{shannon1949communication}, attackers can comprehend the protection method except for the secret keys. In our HCI-based DNN weight protection method, the aging status of the cells in SRAM arrays can be regarded as the secret key. For hardware manufacturers, it is reasonable to assume that they only open the I/O interface of accelerators after they are manufactured and tested. As HCI-induced aging does not cause bit flips, attackers cannot determine the aging status of each cell through the I/O interface. Therefore, the only attack method is the exhaustive search to find which cells are aged and how much the aging degree is. Obviously, the crack complexity is at least $O(2^{|SRAM|}\times |D|)$, where $|SRAM|$ is the total number of SRAM cells in the accelerator and $|D|$ is the number of possible aging degree values. For any existing computer, executing the attack is impossible. If an accelerator has 1Mb SRAM cells, and even if we assume that attackers know the aging degree, namely, $|D|=1$, the attack needs $2^{1048576}$ attempts to find out the aging status, which is clearly an astronomical number.

\section{Conclusion}

Aged chips were once seen as a drawback, but when used with DNNs fine-tuned through the proposed DOFT method, they can achieve remarkable accuracy. This paper skillfully employs HCI to create a hardware-software co-design for DNN weight protection, introducing no additional hardware overhead during DNN inference procedure. We believe this research will inspire the scientific community to concentrate on low-cost model protection, which has greater potential for practical application.
\section*{Acknowledgment}
This work was supported by National Key R\&D Program of China (No. 2023YFB2806300), Beijing Natural Science Foundation Key Program (Z210006), Shenzhen Science and Technology Innovation Commission (SGDX20220530111405040),  
NSFC (No. 62122076, 62122004), RGC GRF (No. 27206321, 17205922, 17212923), Key Research Program of Frontier Sciences, CAS (No. ZDBS-LY-JSC012), CityU SGP (9380132) and ITF MSRP (ITS/018/22MS).

\bibliographystyle{ACM-Reference-Format}
\bibliography{ref}

\end{document}